\documentclass[aps,prx,reprint,preprintnumbers,superscriptaddress,nofootinbib,longbibliography,floatfix]{revtex4-2}
\pdfoutput=1
\usepackage{rotating}
\usepackage{array}
\usepackage{amsmath}
\usepackage[normalem]{ulem}
\usepackage{slashed}
\usepackage{booktabs}
\usepackage[pdftex,table]{xcolor}
\usepackage{xfrac}
\usepackage{xspace}
\usepackage{mathtools}
\usepackage{empheq}
\usepackage{multirow}
\usepackage{amssymb}
\usepackage{url}
\usepackage{comment}
\usepackage{physics}
\usepackage{color,soul}
\usepackage{bbm}
\usepackage[caption=false]{subfig}
\usepackage{graphicx}
\usepackage{adjustbox}
\usepackage{glossaries}
\usepackage{siunitx}

\usepackage{hyperref}
\hypersetup{
  colorlinks=true,
  citecolor=blue,
  linkcolor=blue,
  urlcolor=blue
}
\usepackage{cleveref}

\newcommand{\SUSYSignal}{SUSY150\xspace}
\newcommand{\XSHSignal}{X$\rightarrow$SH\xspace}
\newcommand{\mgg}{\ensuremath{m_{\gamma\gamma}}\xspace}

\newacronym{ggF}{ggF}{gluon--gluon fusion}
\newacronym{VBF}{VBF}{vector boson fusion}
\newacronym{SR}{SR}{Signal Region}
\newacronym{SB}{SB}{Sideband Region}
\newacronym{LHC}{LHC}{Large Hadron Collider}
\newacronym{SM}{SM}{Standard Model}
\newacronym{BSM}{BSM}{Beyond the Standard Model}
\newacronym{AD}{AD}{Anomaly Detection}
\newacronym{IAD}{IAD}{Ideal Anomaly Detector}
\newacronym{CWOLA}{CWoLa}{Classification Without Labels}
\newacronym{CATHODE}{CATHODE}{Conditional Anomaly Detection with Histogram-based Density Estimation}
\newacronym{SUSY}{SUSY}{Supersymmetry}
\newacronym{QCD}{QCD}{quantum chromodynamics}
\newacronym{VAE}{VAE}{Variational Autoencoder}
\newacronym{NF}{NF}{Normalizing Flow}
\newacronym{SIC}{SIC}{Significance Improvement Characteristic}
\newacronym{AUC}{AUC}{area under the curve}
\newacronym{ROC}{ROC}{Receiver Operating Characteristic}
\newacronym{TPR}{TPR}{True Positive Rate}
\newacronym{BDT}{BDT}{Boosted Decision Tree}
\newacronym{BCE}{BCE}{Binary Cross-Entropy}
\newacronym{ML}{ML}{machine learning}

\begin{document}

\title{Weakly Supervised Anomaly Detection\\in Events with a Higgs Boson and Exotic Physics}

\author{Chi Lung Cheng}
\email{ccheng84@wisc.edu}
\affiliation{Physics Division, Lawrence Berkeley National Laboratory, Berkeley, CA 94720, USA}
\affiliation{Department of Physics, University of Wisconsin, Madison, WI 53706, USA}

\author{Sarah Demers}
\email{sarah.demers@yale.edu}
\affiliation{Department of Physics, Yale University, New Haven, CT 06511, USA}

\author{Sascha Diefenbacher}
\email{sdiefenbacher@lbl.gov}
\affiliation{Physics Division, Lawrence Berkeley National Laboratory, Berkeley, CA 94720, USA}

\author{Runze Li}
\email{runze.li@yale.edu}
\affiliation{Department of Physics, Yale University, New Haven, CT 06511, USA}

\author{Benjamin Nachman}
\email{nachman@stanford.edu}
\affiliation{Department of Particle Physics and Astrophysics, Stanford University, Stanford, CA 94305, USA}
\affiliation{Fundamental Physics Directorate, SLAC National Accelerator Laboratory, Menlo Park, CA 94025, USA}

\author{Dennis Noll}
\email{nollde@stanford.edu}
\affiliation{Physics Division, Lawrence Berkeley National Laboratory, Berkeley, CA 94720, USA}
\affiliation{Fundamental Physics Directorate, SLAC National Accelerator Laboratory, Menlo Park, CA 94025, USA}

\begin{abstract}
We present a machine learning-based anomaly detection strategy designed to identify anomalous physics in events containing resonant Standard Model physics and demonstrate this method on the final state of a Higgs boson decaying to two photons.
The demonstration targets high-dimensional deviations in the region of phase space containing the Higgs mass peak in a fully signal-agnostic manner.
A latent-space embedding, learned from event kinematics, enables the use of a large set of potentially sensitive features.
Backgrounds are estimated using a hybrid approach that combines machine learning-based generative modelling with traditional simulation, and a discriminator is trained in the latent space to distinguish data from background estimates.
After applying a selection on the classifier output, the invariant mass distribution of the diphoton system is examined for localized excesses above the simulated Higgs peak.
We benchmark the sensitivity of this strategy using simplified simulated proton--proton collisions corresponding to data recorded during Run 2 of the LHC, and show that the method can provide significant improvements in sensitivity, even for small signal injections that could remain undetected in an inclusive analysis.
These results demonstrate that the proposed strategy is a promising and viable approach for future searches and should be applied to recorded collider data.
\end{abstract}

\maketitle
\flushbottom


\section{Introduction}
\label{sec:intro}
The \gls{LHC} has enabled the exploration of the \gls{SM} with unprecedented precision.
Nevertheless, there remain compelling motivations to search for physics \gls{BSM}, such as the existence of dark matter~\cite{Bertone:2004pz}, the observed matter-antimatter asymmetry, or the hierarchy problem~\cite{Arkani-Hamed:1998jmv}.
A major challenge in searches for \gls{BSM} physics is the vast number of possible models, which far exceeds the number of feasible targeted search efforts.

\gls{AD}~\cite{Kasieczka:2021xcg,Karagiorgi:2021ngt,Aarrestad:2021oeb,Belis:2023mqs} offers a complementary search paradigm, enabling the study of signatures that correspond to a wide range of \gls{BSM} scenarios simultaneously.
This approach is designed to complement dedicated, model-specific searches by providing sensitivity across broad model spaces.
\Gls{AD} searches have seen use at both the ATLAS~\cite{ATLAS:2023azi, ATLAS:2023ixc, ATLAS:2020iwa, ATLAS:2025obc} and CMS~\cite{CMS:2024nsz} experiments. 
Different \gls{AD} methods can be categorized in part on the prior knowledge of the signal.
In this work, we present a weakly supervised \gls{AD} analysis strategy, following the \gls{CWOLA}~\cite{Metodiev:2017vrx,Collins:2018epr,Collins:2019jip} and \gls{CATHODE}~\cite{Hallin:2021wme} approaches.
This method can achieve close-to-optimal sensitivity depending on the signals assumed to be present in the data.

Our new \gls{AD} strategy is aimed at identifying anomalies associated with an \gls{SM} resonance.
This extends \gls{CATHODE} to the case where the background has two components: a non-resonant part estimated from data sidebands (as in standard \gls{CATHODE}) and a resonant part estimated from simulation or alternative methods.
As a demonstrator, we apply it to final states containing a Higgs boson $H$ in association with any other particles $X$, a setup we denote as \textsc{HAXAD} (\textbf{H}iggs and \textbf{X} \textbf{A}nomaly \textbf{D}etection).

The Higgs boson plays a unique role in the \gls{SM} as the only elementary scalar particle and poses a potential portal to new physics.
The presence of \gls{BSM} effects in this sector could alter the inclusive and differential production cross sections of the Higgs boson, modify its decay rates, or introduce entirely new production mechanisms.
For instance, the Higgs boson could be produced in final states from \gls{SUSY} cascades~\cite{Golfand:1971iw,Volkov:1973ix,Wess:1974tw,Wess:1974jb,Ferrara:1974pu,Salam:1974ig}, exotic top-quark decays~\cite{Guasch:1999jp,Bejar:2000ub,Eilam:2001dh,Aguilar-Saavedra:2002phh,Cao:2007dk}, or via vector-like quarks~\cite{delAguila:1982fs,Aguilar-Saavedra:2009xmz}, as motivated in~\cite{ATLAS:2023omk}.
To date, no significant deviations from the \gls{SM} have been observed in the Higgs sector -- but most analyses have been tailored to specific signal models, offering high sensitivity only within limited regions of parameter space.

Within the Higgs sector, our search focuses on the $H \rightarrow \gamma\gamma$ decay channel.
While the branching ratio for $H \rightarrow \gamma\gamma$ is relatively small, this channel is highly competitive due to its excellent diphoton mass resolution, strong background rejection, and the ability to model the dominant background processes with analytic functions.

In this work, we demonstrate our proposed strategy on simplified simulated datasets and aim to illustrate how the method could be applied to real data from experiments such as ATLAS~\cite{ATLAS:2008xda} or CMS~\cite{CMS:2008xjf}.
A classical \gls{AD} analysis was previously performed by ATLAS in the $H \rightarrow \gamma\gamma$ channel~\cite{ATLAS:2023omk}, following the approach of earlier searches~\cite{ATLAS:2018zdn,CMS:2020zjg} that look for excesses across a large number of exclusive final states, none of which are optimized for a specific signal model.
In contrast, the present strategy incorporates \gls{ML} techniques and is optimized directly on the target data, enabling it to adjust to the characteristics of any potential new physics signal.
Additionally, while the new strategy is demonstrated here for the $H \rightarrow \gamma\gamma$ decay channel, the proposed strategy is applicable to any known resonance and is also not restricted to collider-based experiments.

The paper is organized as follows:
\Cref{sec:datasets} describes the used dataset, including the event pre-selection and reconstruction of event features.
\Cref{sec:strategy} explains the analysis strategy, and \cref{sec:results} presents the results of the demonstrative analysis.
Finally, \cref{sec:conclusion} provides our conclusions and outlook.

\section{Datasets}
\label{sec:datasets}
The analysis uses simulated data corresponding to proton--proton collision at a center-of-mass energy of $\sqrt{s} = \SI{13}{\TeV}$ with an integrated luminosity of $\mathcal{L} = \SI{137}{\per\femto\barn}$.
The event generation is performed with \textsc{MadGraph5\_aMC@NLO}~v3.5.9~\cite{Alwall:2011uj} and \textsc{Pythia}~v8.312~\cite{Sjostrand:2006za,Sjostrand:2014zea}.
All processes are generated at leading order in \gls{QCD} and do not include contributions from additional simultaneous proton--proton interactions (pileup) or multi-parton interactions.
A Higgs boson mass of $m_H = \SI{125}{\GeV}$ is used for all processes involving Higgs bosons, and all Higgs decays are simulated at the parton-shower level.
A simplified detector simulation is performed using \textsc{Delphes}~v3.5.0~\cite{deFavereau:2013fsa,Mertens:2015kba} with a configuration corresponding to a simplified ATLAS detector layout.
A link to the repository containing the full simulation framework is given at the end of the paper.
The use of leading order approximations and the simplified detector simulation will result in a less accurate modeling than using higher order calculations and a dedicated full detector simulation; however, we do not expect a more accurate simulation to change the quantitative results. 
Likewise, any classical benchmark results may differ from ones obtained using full simulation.
Therefore, we do not aim to quantify the performance of any classical benchmark.
Instead, we use them as a relative comparison point.

The \gls{SM} background is composed of two main categories: the non-resonant continuum and the resonant Higgs backgrounds.
The continuum background includes non-resonant processes producing two prompt photons and up to three jets in the hard process.
To regulate divergences in the generation, minimal phase-space cuts are applied to the transverse momentum of the photons, requiring $p_{T} > \SI{32}{\GeV}$.

The resonant background includes the production of a \gls{SM} Higgs boson with its subsequent decay into $H \rightarrow \gamma\gamma$.
It includes the four leading Higgs production channels, \gls{ggF}, \gls{VBF}, associated production with a weak boson ($VH$, with $V = W, Z$), production of the Higgs boson together with a top-quark--antiquark pair ($t\bar{t}H$), and other subleading production processes.

Two classes of \gls{BSM} signal processes are considered to benchmark the sensitivity of the \emph{signal-agnostic} analysis: \gls{SUSY} signals and signals from an extended Higgs sector.  
The \gls{SUSY} signals involve the production of a chargino ($\tilde{\chi}_1^\pm$) in association with the next-to-lightest neutralino ($\tilde{\chi}_2^0$), with prompt decays  
$\tilde{\chi}_1^\pm \rightarrow W^\pm \tilde{\chi}_1^0$ and $\tilde{\chi}_2^0 \rightarrow H \tilde{\chi}_1^0$,  
where $W^\pm$ is a \gls{SM} W boson, $H$ is the \gls{SM} Higgs boson, and $\tilde{\chi}_1^0$ is the lightest neutralino~\cite{Fayet:1976et,Fayet:1977yc,Farrar:1978xj,Alwall:2008ag,LHCNewPhysicsWorkingGroup:2011mji,ATLAS:2020qlk}.
For the benchmark signal, denoted as \SUSYSignal, the masses of the chargino and the next-to-lightest neutralino are set to $m_{\tilde{\chi}_1^\pm} = m_{\tilde{\chi}_2^0} = \SI{150}{\GeV}$ and the lightest neutralino mass is set to $m_{\tilde{\chi}_1^0} = \SI{0.5}{\GeV}$.
The extended Higgs sector signals involve a heavy \gls{BSM} Higgs boson ($X$) which decays to a lighter \gls{BSM} Higgs boson ($S$) and the \gls{SM} Higgs boson ($H$)~\cite{Robens:2019kga,Basler:2018dac,Baum:2019pqc,CMS:2022suh,CMS:2021yci,CMS:2023boe,ATLAS:2023tkl}.
For the benchmark signal, the mass of $X$ is set to \SI{750}{\GeV}, and the mass of $S$ is set to \SI{100}{\GeV}.  
The heavy \gls{BSM} Higgs boson is constrained to exclusively decay as \XSHSignal, and the lighter \gls{BSM} Higgs boson is constrained to decay to light quarks only, to make its missing transverse energy ($E_T^{\text{miss}}$) distribution significantly different from the \gls{SUSY} signals.
Only $H \rightarrow \gamma\gamma$ decays are realized for the Higgs boson in all signal processes.  

The simulated events are used to construct datasets corresponding to the targeted integrated luminosity.
Events are randomly sampled from all simulated events according to weights derived from their production cross section and generator-level event weights, ensuring that the final datasets correspond to the expected number of recorded events.
To test different signal hypotheses, signal events are injected into the background at various levels of signal significance $s = S / \sqrt{B}$ with the expected inclusive number of signal ($S$) and background ($B$) events in the respective signal region.

The event selection broadly aligns with the standard event selection of $H \rightarrow \gamma\gamma$ analyses from the ATLAS collaboration.
As our primary analysis objects, we select the two highest transverse momentum ($p_T$) photons with an invariant mass \mgg in the range of $105 < \mgg < \SI{160}{\GeV}$.
The leading and sub-leading photons are required to have a minimum $p_T$ of \SI{35}{\GeV} and \SI{25}{\GeV}, respectively, to ensure the diphoton trigger is fully efficient and to reduce potential turn-on effects.
To further enhance the selection of photons from a resonant decay, the leading and sub-leading photons must satisfy $p_T/\mgg > 0.4$ and $p_T/\mgg > 0.3$, respectively.
To suppress backgrounds from jets misidentified as photons, both photons are required to be isolated by ensuring the sum of charged-particle transverse momenta in a cone of $\Delta R = 0.5$ around the photon is less than 50\% of its $p_T$.
A similar isolation criterion is applied to leptons.
Jets are reconstructed using the anti-$k_T$ algorithm~\cite{Cacciari:2005hq, Cacciari:2011ma, Cacciari:2008gp} with $R=0.4$.
A threshold of $p_T > \SI{20}{\GeV}$ is applied to reduce contributions from pileup and underlying event.

To preserve the model-agnostic nature of the analysis, each event is characterized by nine high-level kinematic features.
These variables are chosen for their general sensitivity to a broad range of BSM physics processes, rather than being optimized for any specific signal signature.
The features comprise the diphoton transverse momentum ($p_T^{\gamma\gamma}$) and angular separation ($\Delta R_{\gamma\gamma}$); the transverse momentum of the two leading jets ($p_T^{J_1}$, $p_T^{J_2}$), their invariant mass ($m_{JJ}$), and their angular separation ($\Delta R_{JJ}$); a flag indicating the presence of one or more isolated leptons; the scalar sum of jet transverse momentum, $H_T$, for jets with $p_T > \SI{30}{\GeV}$; and the missing transverse energy, $E_T^{\text{Miss}}$.
If an event contains fewer than two jets, the kinematic variables for any non-existent jets, as well as dijet features, are assigned a default value of zero.
There are many more features that characterize events and are potentially sensitive to \gls{BSM} physics and the use of an even more extended set of features is left for future work.

\section{Analysis Strategy}
\label{sec:strategy}

In this analysis, we employ an \gls{ML}-based \gls{AD} strategy to search for discrepancies between observed data and estimated background events.
The background estimation is performed using a hybrid approach that combines \gls{ML} with traditional simulation-based techniques.
Classifiers, trained to distinguish between data and background estimation, operate in a high-dimensional embedded feature space, enabling the use of a large number of variables that are potentially sensitive to \gls{BSM} physics.
The final sensitivity is extracted with a bump hunt in the invariant mass distribution of the diphoton system.

Similar to other \gls{CATHODE}-like \gls{AD} approaches, we need to define a \gls{SR} in which we search for a potential signal and a \gls{SB} which we assume to be signal-depleted.
As we search for anomalies related to the Higgs boson, the \gls{SR} is constrained to be centered around the Higgs mass. 
Therefore, our \gls{SR} is fixed and does not need to be moved -- a new feature of our approach that is anchored on the particular resonance.
For this analysis, we chose a \gls{SR} width of \SI{10}{\GeV}, centered around the approximate Higgs mass, resulting in \gls{SR} boundaries of $[\SI{120}{\GeV}, \SI{130}{\GeV}]$.

The background estimation in the \gls{SR} is complicated by unphysical artefacts in the data, such as hard cutoffs and default placeholder values.  
To mitigate these effects, we encode our data into a latent space using a \gls{VAE}~\cite{kingma2014autoencoding}, which smooths over such artefacts while preserving physics-relevant information.  
The two key innovations of this analysis are the use of a high-dimensional embedding space for better sensitivity and the hybrid background estimation method, enabling a completely new analysis scope.

All steps of the analysis -- including the latent space embedding, background estimation, signal classification, and statistical inference -- are described in the following sections.

\subsection{Latent space embedding}
\begin{figure}
\centering
    \centering
    \includegraphics[page=1, width=\linewidth]{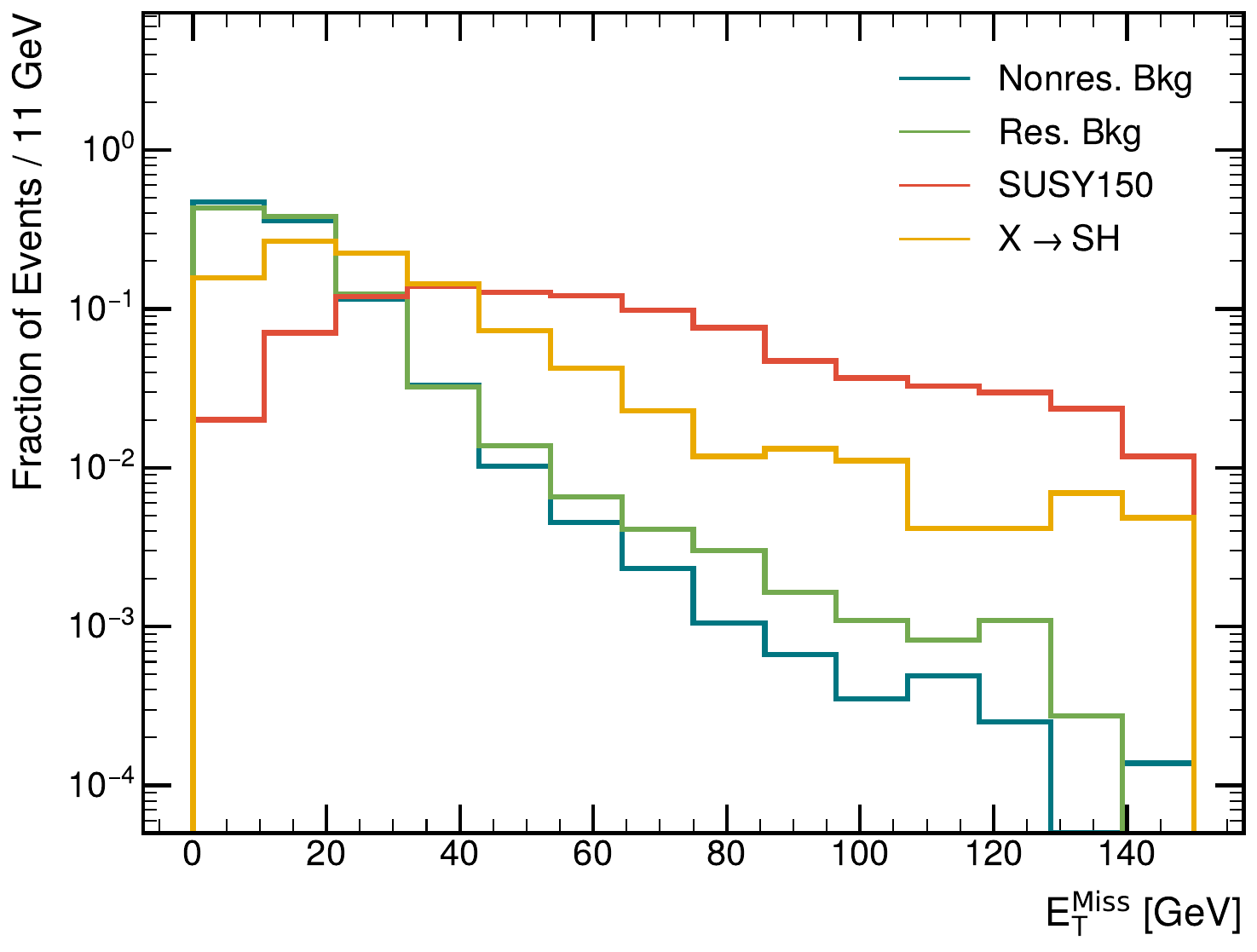}
    \caption{
        Distribution of the missing transverse momentum $E_T^{\text{Miss}}$ for the non-resonant $\gamma\gamma$ + jets background, the \gls{SM} Higgs background, and the benchmark signals \SUSYSignal and \XSHSignal.
    }
    \label{fig:met}
\end{figure}

The nine selected kinematic features from the \gls{SB} data are pre-processed by applying a logarithmic transformation to the $p_T$ and mass features given in units of \si{\GeV}.  
All features are then standardized to have zero mean and unit variance.
The original distributions of the missing transverse momentum $\textrm{E}^{\textrm{Miss}}_{\textrm{T}}$ in the \gls{SR} are shown in \cref{fig:met} separately for the non-resonant $\gamma\gamma$ + jets background, the resonant \gls{SM} Higgs background, and the benchmark signals \SUSYSignal and \XSHSignal.

To remain as general as possible, we do not impose a jet multiplicity requirement.  
As a result, some events lack defined values for jet-related observables such as $m_{JJ}$, $\Delta R_{J_1 J_2}$, $p_{T,J_1}$, and $p_{T,J_2}$.  
For such cases, these observables are set to zero, corresponding to the mean of their distributions after pre-processing.  
This introduces large peaks at singular values in the respective feature distributions.  
Generative models, being inherently probabilistic, struggle to reproduce such discrete spikes, which complicates the estimation of a continuous \gls{SR} background.
This can degrade the performance of downstream weakly supervised classifiers, reducing signal sensitivity or even causing false discoveries due to background sculpting.

To address this, we employ a \gls{VAE}~\cite{kingma2014autoencoding} to encode each event into a lower-dimensional latent representation.  
The \gls{VAE} smooths over unphysical artefacts while preserving physics-relevant information for distinguishing signal from background.
Therefore, the \gls{VAE} latent space provides a simpler but still effective feature space for generative modelling.

The \gls{VAE} is trained on \gls{SB} data, mapping the nine selected input features into a four-dimensional latent space.  
The model is implemented in \textsc{PyTorch}~\cite{DBLP:journals/corr/abs-1912-01703}.
Its architecture consists of an encoder and decoder, each with three hidden layers of 128 nodes.  
The training uses the \textsc{Adam} optimizer~\cite{adam} with an initial learning rate of 0.001 and a cosine annealing scheduler.  
The \gls{SB} dataset is randomly split into 80\% training and 20\% validation subsets.  
The maximum number of epochs is 200, with early stopping if the validation loss does not improve for 20 epochs.

Once trained, the VAE encoder maps both \gls{SR} and \gls{SB} events into the latent space.  
The latent feature distributions for the events in the \gls{SR} are shown in \cref{fig:gen_features_comp}.  
Compared to the raw physical features, the encoded features are significantly smoother and have no singular peaks from default values.
In the \gls{SR}, both benchmark signals exhibit distinct latent-space distributions compared to the $\gamma\gamma$+jets and \gls{SM} Higgs backgrounds, indicating that the \gls{VAE} effectively captures and represents the underlying physics.

Our embedding approach is comparable to the Latent-\gls{CATHODE} concept~\cite{Hallin:2022eoq}.
However, it differs since we use a \gls{VAE} for the embedding, rather than an \gls{NF}, and we use a generative model trained on the embedding space, rather than approximating it with a Gaussian distribution, as explained in the next section.

\subsection{Background Estimation}

When estimating the \gls{SR} events, we need to distinguish between two contributions: (i) di-photon events produced in nonresonant electromagnetic processes, and (ii) di-photon events produced by decays of the \gls{SM} Higgs boson. 
Both contributions require separate treatment.

The nonresonant part is expected to be continuous between \gls{SR} and \gls{SB}. 
As a result, we used a generative model, conditioned on the diphoton mass and trained on the \gls{SB}, to interpolate into the \gls{SR}. 
This approach is purely data-driven and does not rely on simulation.

The generative model we use is a \gls{NF}~\cite{papamakarios2021normalizingflowsprobabilisticmodeling}. 
Our \gls{NF} is implemented in \textsc{PyTorch}~\cite{DBLP:journals/corr/abs-1912-01703} using the nFlows package~\cite{nflows}.
The model itself consists of 6 Rational Quadratic Spline (RQS)~\cite{durkan2019neural} Transformation Layers. 
Each transformation layer has a backend consisting of a 2-block residual network with a hidden layer width of $8 \times \# \text{features} = 72$ nodes. 
Interspersed between each RQS layer is a random permutation layer.
The \gls{NF} is trained on the data points encoded by the \gls{VAE} model. 
As this is an already well-regularized representation, only minor mean and width scaling is required for the \gls{NF} training. 

For the training, the full \gls{SB} data is split into train and validation sets, with the training set comprising 80\% of the data.
The \gls{NF} training uses an \textsc{Adam}~\cite{adam} optimizer with an initial learning rate of $0.001$. 
We use an early stopping procedure that monitors the validation loss and halves the learning rate every time the validation loss does not improve for $2$ epochs, with an epoch being defined as $1000$ iterations.
After the learning rate has been reduced below $10^{-8}$, the training terminates. 
To minimize the impact of random model initialization and batch ordering, we train an ensemble of 4 \gls{NF} models, each with a different train and validation split. 

Several alternative approaches were tried to improve the performance and accuracy of the generative model. 
Notably, we investigated the use of Conditional Flow Matching (CFM) models~\cite{lipman2023flowmatchinggenerativemodeling} as well as the use of a classifier network to reweight the generated samples~\cite{Diefenbacher:2020rna}.
However, none of the attempted approaches led to a performance improvement over the \gls{NF} model, and in some instances, the alternate models were found to be more prone to background sculpting.

After the model is trained, we use it to generate \gls{VAE}-latent space events in the \gls{SR}. 
To this end, we need to sample diphoton masses from the \gls{SR}. 
We achieve this by fitting an exponential falling function to the diphton mass spectrum in the \gls{SB}.
The fitted function has the form $a  e^{-x \cdot c}$ with the normalized diphoton mass $x$ and the free parameters $a$ and $c$.
The conditional inputs to our model are then sampled from the fitted function in the \gls{SR} using rejection sampling. 
Each \gls{NF} in the ensemble is used to sample 250,000 events, which are combined into one generated set of 1 million events.
This oversampling has been shown to improve the performance of the weakly supervised classifiers.
The generated events are then weighted such that the sum of all generated event weights corresponds to the number of expected events we obtain by integrating the fitted diphoton mass function in the \gls{SR}.

The resonant contribution from the \gls{SM} Higgs is only present in the \gls{SR} and can therefore not be modeled with an ML-based and data-driven approach from the \gls{SB}, like the non-resonant background.
Instead, we estimate the contribution of the Higgs process directly from classical simulations as described in \cref{sec:datasets}.
In this demonstrator, the resonant background is modeled with different events from the same simulations in the pseudo-data and the simulated data -- to model the effect of potential differences which might occur in an analysis on recorded data, a 5\% normalization uncertainty is added on the total cross section of the simulated Higgs process as described in \cref{sec:inference}.

\begin{figure*}
    \captionsetup[subfigure]{labelformat=empty}
    \centering
    \subfloat[]{
        \centering
        \includegraphics[page=1, width=0.45\linewidth]{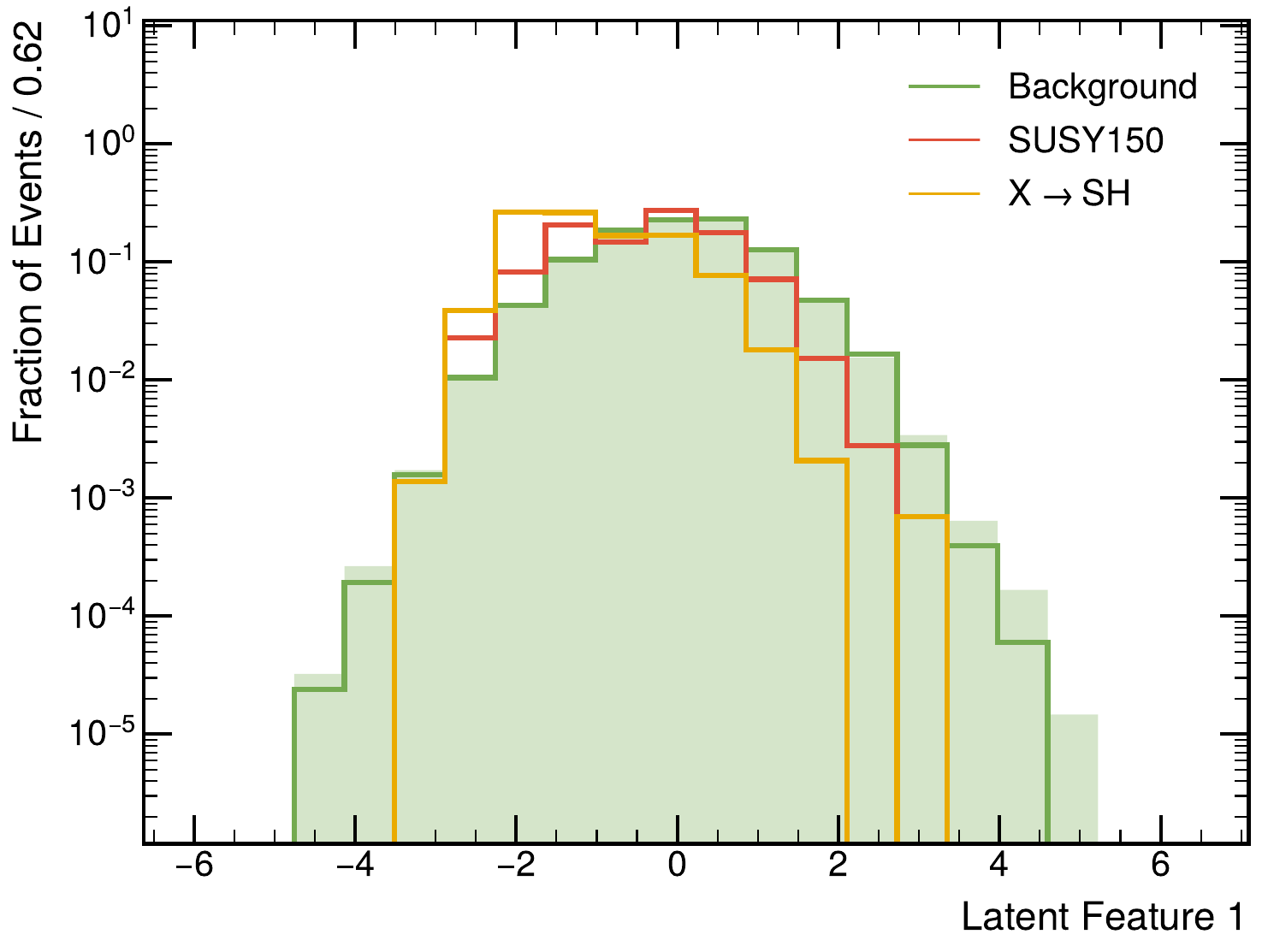}
    }%
    \hfill
    \subfloat[]{
        \centering
        \includegraphics[page=2, width=0.45\linewidth]{figures/Background_estimation/sr_data_gen_plot_embeded}
    }\\%
    \subfloat[]{
        \centering
        \includegraphics[page=3, width=0.45\linewidth]{figures/Background_estimation/sr_data_gen_plot_embeded}
    }%
    \hfill
    \subfloat[]{
        \centering
        \includegraphics[page=4, width=0.45\linewidth]{figures/Background_estimation/sr_data_gen_plot_embeded}
    }\\
    \caption{
      Comparison between the distributions in the 4-dimensional latent space for events in the \gls{SR}.
      The shapes of the background in the data (line) and the estimated background (filled) indicate a high accuracy of the background estimation.
      The shapes of the \SUSYSignal and \XSHSignal signals indicate the separation from the backgrounds in the latent space.
    }
    \label{fig:gen_features_comp}
\end{figure*}

\Cref{fig:gen_features_comp} shows a comparison of the distributions of pseudo-data and background estimation in the \gls{VAE} space.
It can be seen that the ML technique can model the distribution of the pseudo-data.
The estimation of the resonant background agrees with the data by construction in this demonstrator analysis, as previously described.
Classifiers trained to distinguish between the pseudo-data without any injected signal and the background estimation achieve an \gls{AUC}s close to 0.5. 
This value is consistent with random guessing, meaning the classifier cannot separate the data from generated samples, indicating that the background is accurately modelled.

\subsection{Signal Classification}
\label{sec:classifier}

After generating the background reference sample, weakly supervised classifiers are trained to distinguish data from the background reference.
As other recent \gls{AD} strategies~\cite{Finke:2023ltw,Freytsis:2023cjr}, we employ \gls{BDT} classifiers for this task.
The classifiers are trained with a \gls{BCE} loss and implemented using the XGBoost package~\cite{DBLP:journals/corr/ChenG16}.
We observed that reducing the \gls{BDT} size helps prevent overfitting to small discrepancies between the generated background samples and the data.
This avoids background sculpting and improves signal sensitivity.
Consequently, the maximum number of estimators is set to 50, and the maximum tree depth is limited to 3.
A learning rate of 0.01 is used, and early stopping is applied with a patience of 5 epochs.

We further employ a 5-fold cross-validation strategy to prevent overfitting and use our data as efficiently as possible. 
We randomly split our data set into five folds, and for each fold, a \gls{BDT} is trained using three folds as the training set, one fold as the validation set, and one fold as the test set.
After training, each \gls{BDT} is applied to its corresponding test set, and the five test sets are then merged to reconstruct the full dataset.
As a result, every data event receives a classifier score from a \gls{BDT} that was not trained using that event.
To reduce fluctuations from limited training statistics and model initialization, we use the mean of ensembles of classifiers.
Each classifier ensemble consists of four \glspl{BDT}, each trained using the 5-fold cross-validation, with different data splits and background samples generated from different \gls{VAE}-generative model pairs.

\begin{figure*}
    \centering
    \subfloat[]{\label{fig:classifier:SUSY150}
        \centering
        \includegraphics[width=0.45\linewidth]{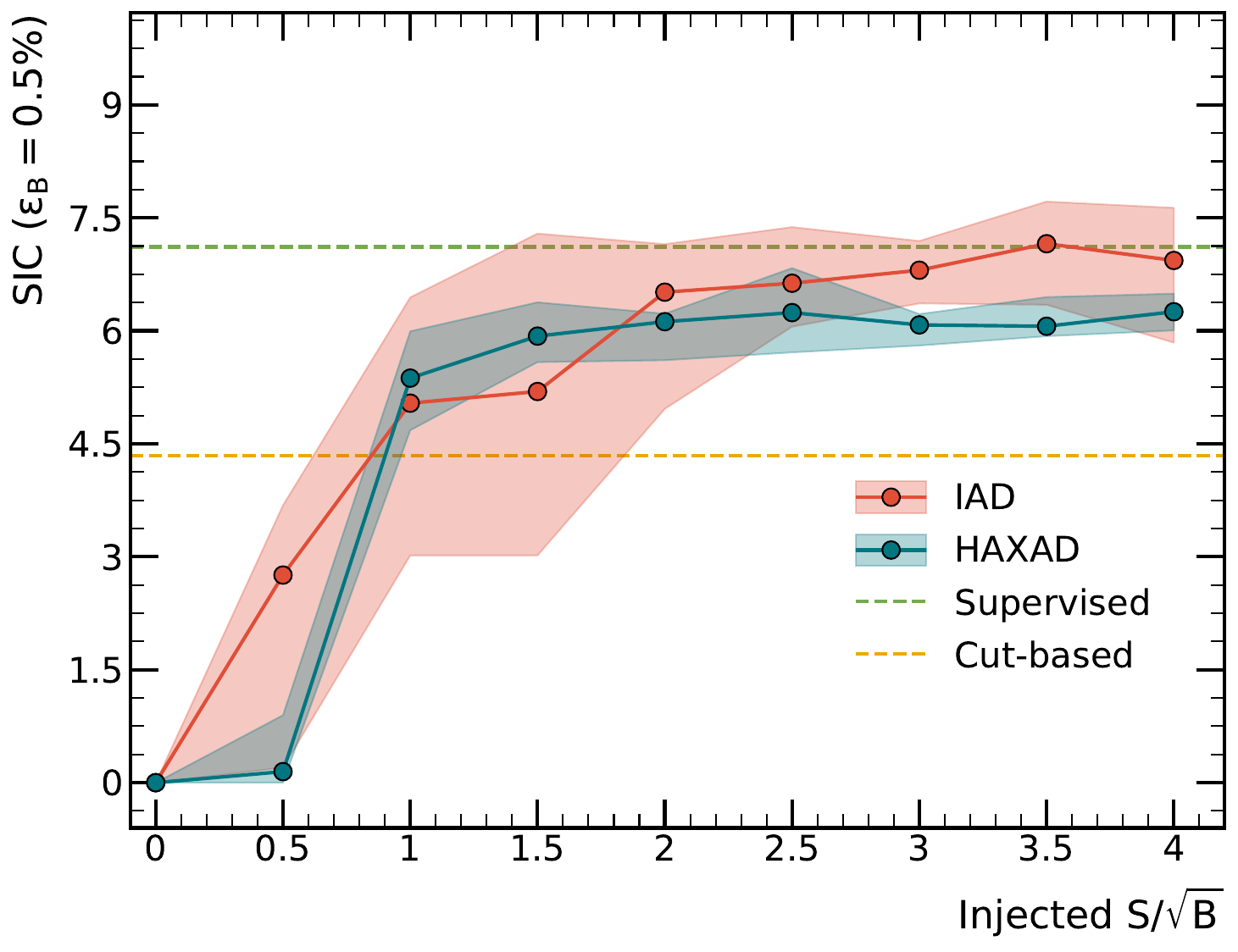}
    }%
    \hfill
    \subfloat[]{\label{fig:classifier:XSH}
        \centering
        \includegraphics[width=0.45\linewidth]{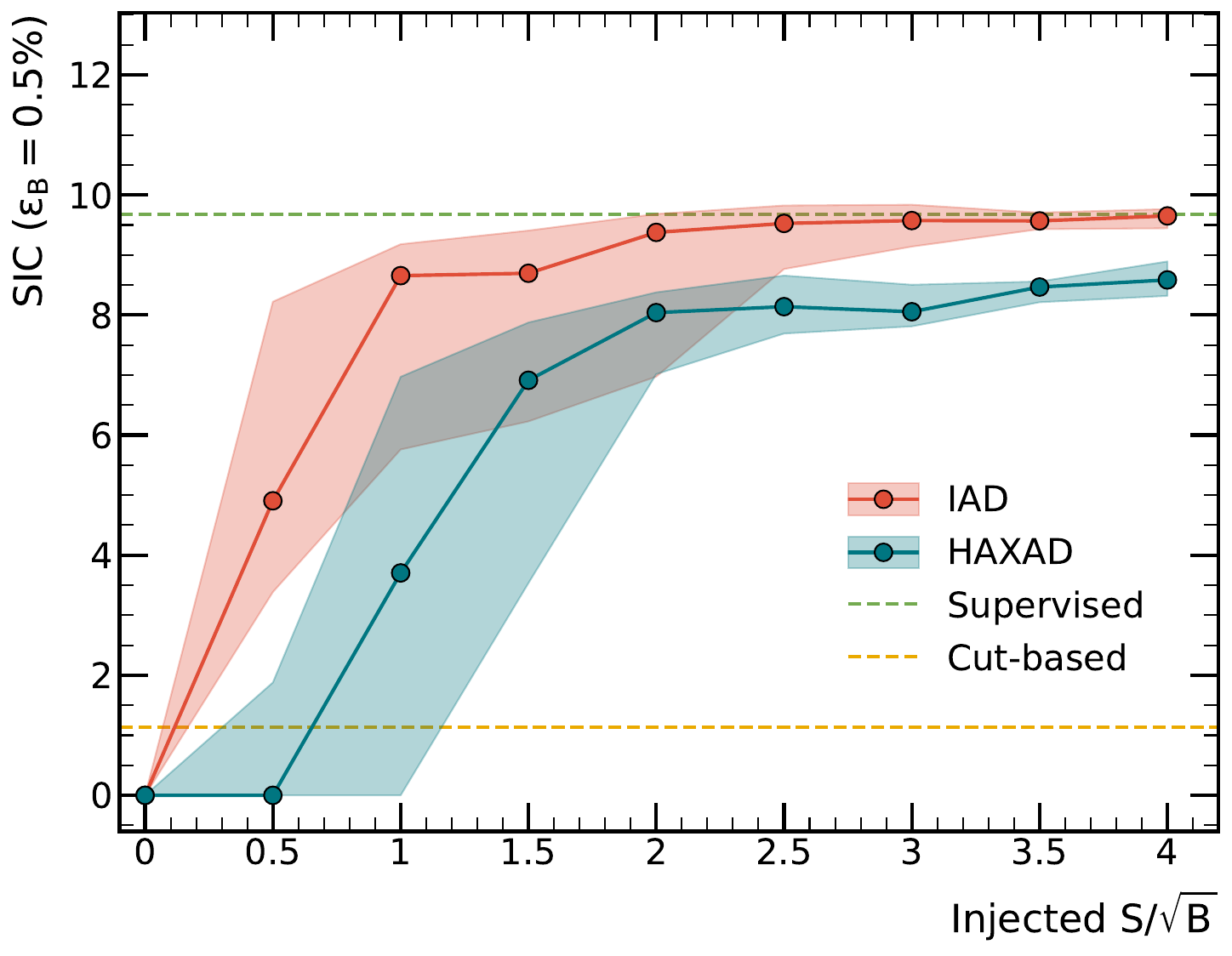}
    }
    \caption{
        \Gls{SIC} from a cut on the classifier output over the respective \gls{TPR} for a) the \SUSYSignal and b) the \XSHSignal benchmark signal.
        The used dataset contains the respective signal with an inclusive pre-classifier significance of $1\sigma$.
        The plot shows the median \gls{SIC} and the $1\sigma$ quantile spread of 50 classifier ensembles, of which each 5 are trained on a different random instance of the training data.
    }
    \label{fig:classifier}
\end{figure*}

The \gls{SIC} between signals and backgrounds in the data is used to evaluate the performance of the classifiers.
\Cref{fig:classifier:SUSY150} (\cref{fig:classifier:XSH}) shows the maximum \gls{SIC} of the respective classifier over the injected amount of the \SUSYSignal (\XSHSignal) benchmark signal.
Shown are the median and $1\sigma$ quantile spread of 50 classifier ensembles, of which each 5 are trained on a different random instance of the initial dataset and a different set of background samples generated from different \gls{VAE} and generative model combinations.
For both tested benchmark models, the significance improvement achieved by the \gls{IAD} classifier approaches that of the fully supervised classifier.  
The HAXAD classifier yields a slightly smaller improvement than the \gls{IAD} classifier, possibly indicating a small information loss in the \gls{VAE} embedding.
Nevertheless, both \gls{IAD} and HAXAD outperform the cut-based selection for a 1\,$\sigma$ signal injection in the \SUSYSignal scenario, where the cut is optimized, and for even smaller signal injections in the \XSHSignal scenario, where the cut is not optimized.

\subsection{Inference}
\label{sec:inference}

The statistical interpretation of the data is performed using a Python implementation of the BumpHunter algorithm~\cite{Vaslin:2022bds}, which searches for localized excesses in the \mgg distribution.
Since the analysis targets anomalies associated with the Higgs boson, the search window is fixed to the \gls{SR} of $120 < \mgg < \SI{130}{\GeV}$, which avoids the look-elsewhere effect from scanning multiple windows.
We perform the bump-hunt on \mgg histograms with a bin width of \SI{0.5}{\GeV}.
To enhance the separation between potential new physics signals and the \gls{SM} background, a tight selection is applied.
To this end, we place a cut on the classifier score such that we retain the 0.5\% most signal-like parts of data in the \gls{SR}.
To mitigate fluctuations from limited training statistics and model initialization of the \glspl{VAE} and generative models, an average from 5 classifier ensembles is used, as explained in the following.
Here, a classifier ensemble refers to the ensemble of four individual classifiers as explained in~\cref{sec:classifier}.
All classifiers of every classifier ensemble are trained with different data splitting and random model initializations.
For each classifier ensemble, we apply the score-cut and build a histogram from the remaining events.
The final histogram for the hypothesis test is then obtained by averaging over the histograms from each classifier ensemble.
The non-resonant background after the score-cut is estimated by fitting the same exponential function as used in the background modelling, $a  e^{-x \cdot c}$ with the normalized diphoton mass $x$ and the free parameters $a$ and $c$, to the post-cut data in the \gls{SB}.
The resulting function is then interpolated into the \gls{SR}.

We quantify the significance of any observed excess using $3.5 \times 10^6$ pseudo-experiments, incorporating both statistical and systematic uncertainties.  
The dominant systematic contributions arise from the classifier ensemble and the \gls{SM} background modelling.  
A total of 100,000 bootstrapped variations of the classifier ensemble are constructed and used in equal proportion across all pseudo-experiments, to limit computational resources.
For each bootstrapped ensemble, the full procedure for deriving the post-selection histograms is repeated to obtain the corresponding varied background predictions.
The statistical uncertainty on the non-resonant background is incorporated by resampling the sideband data before performing the fit.
This results in variations in the fit result, effectively propagating this uncertainty into the background estimate in the \gls{SR}.
To account for theoretical uncertainties in the \gls{SM} Higgs production cross section and decay branching fractions, the normalization of the resonant background is varied according to a log-normal distribution.
For the purposes of this demonstrator, we assume a standard deviation of 5\%. 
A detailed uncertainty calculation of this is beyond the scope of this study.
The final background prediction for each pseudo-experiment is the sum of the non-resonant and resonant components, with an additional Poisson fluctuation applied to simulate statistical variations.
Finally, the compatibility of the observed data with the background-only hypothesis is determined using a frequentist approach:
For each pseudo-experiment, we consider its individual local p-value.
The global p-value is calculated as the fraction of pseudo-experiments with a p-value smaller than that observed in the data.
This represents the probability of observing such a deviation by chance after considering all statistical and systematic uncertainties.

\section{Results}
\label{sec:results}

The HAXAD method is compared against two other methods, the \gls{IAD} and a cut-based approach.
In the \gls{IAD} setup, the background samples are directly drawn from the same simulated samples from which the pseudo data is constructed.
In this case, aside from statistical fluctuations, the only distinguishing factor between the data and the background reference is the presence of signal.
The \gls{IAD} setup also uses the nine physical features instead of the four encoded latent space features to train the \gls{CWOLA} classifiers.
It therefore defines the theoretical upper bound of the performance of the HAXAD setup under the assumption of a perfect background estimation and a latent space embedding that is fully efficient to the assumed signal model.
The workflows following the classifier training are identical to the ones used in the HAXAD method.

The cut-based method is similar to what was done in the categorical ATLAS H+X analysis~\cite{ATLAS:2023omk}, which applies one-dimensional cuts on selected kinematic features to define regions with higher signal fraction.
Instead of multiple cuts, only one cut is employed, which is optimized for the \SUSYSignal but not for the \XSHSignal benchmark signal.
The inference workflow of the cut-based method after the cut is identical to the HAXAD method.

\begin{figure*}
    \centering
    \subfloat[]{\label{fig:spectrum:SUSY150}
        \centering
        \includegraphics[width=0.45\linewidth]{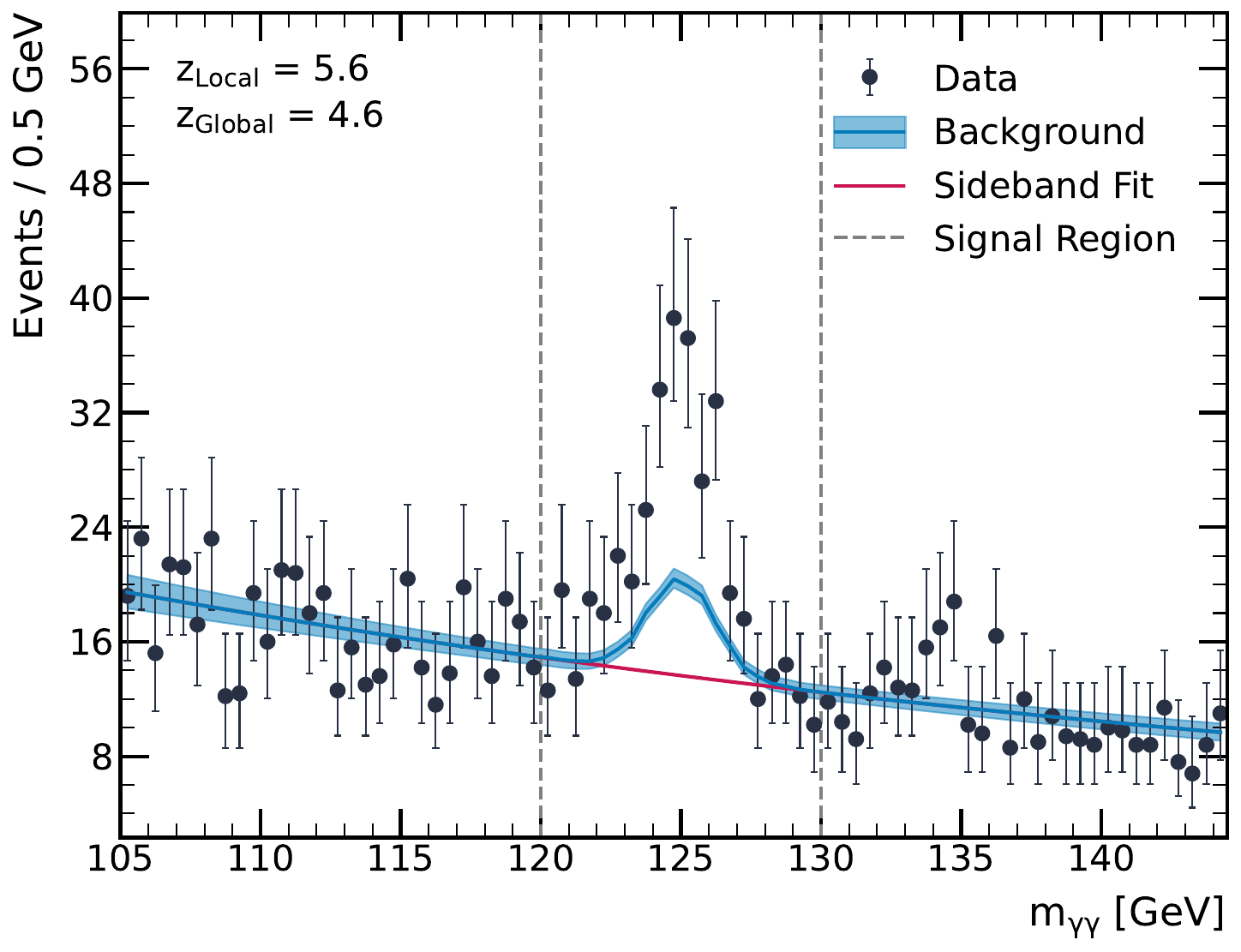}
    }%
    \hfill
    \subfloat[]{\label{fig:spectrum:XSH}
        \centering
        \includegraphics[width=0.45\linewidth]{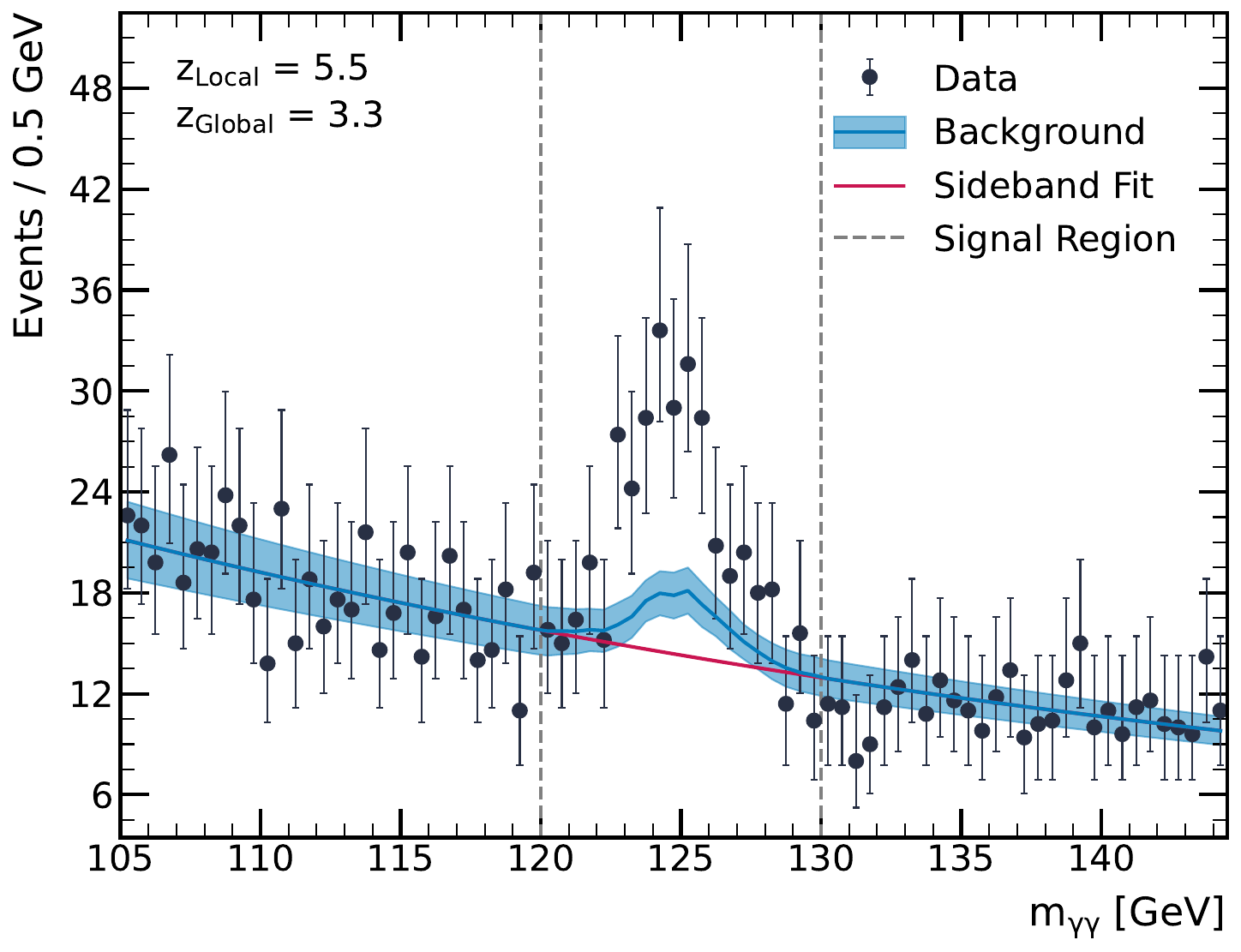}
    }
    \caption{
        \mgg spectrum after the classifier-based selection in the case of an injected signal with an inclusive $1\sigma$ pre-selection significance of a) the \SUSYSignal and b) the \XSHSignal benchmark signal.
    }
    \label{fig:spectrum}
\end{figure*}

The \mgg spectrum after the classifier-based selection, including the bump hunting results, using the \SUSYSignal (\XSHSignal) signal model at 1\,$\sigma$ signal injection significances, is shown in \cref{fig:spectrum:SUSY150} (\cref{fig:spectrum:XSH}), where a global significance of 4.6\,$\sigma$ (3.3\,$\sigma$) is observed.

\begin{figure*}
    \centering
    \subfloat[]{\label{fig:scan:SUSY150}
        \centering
        \includegraphics[width=0.45\linewidth]{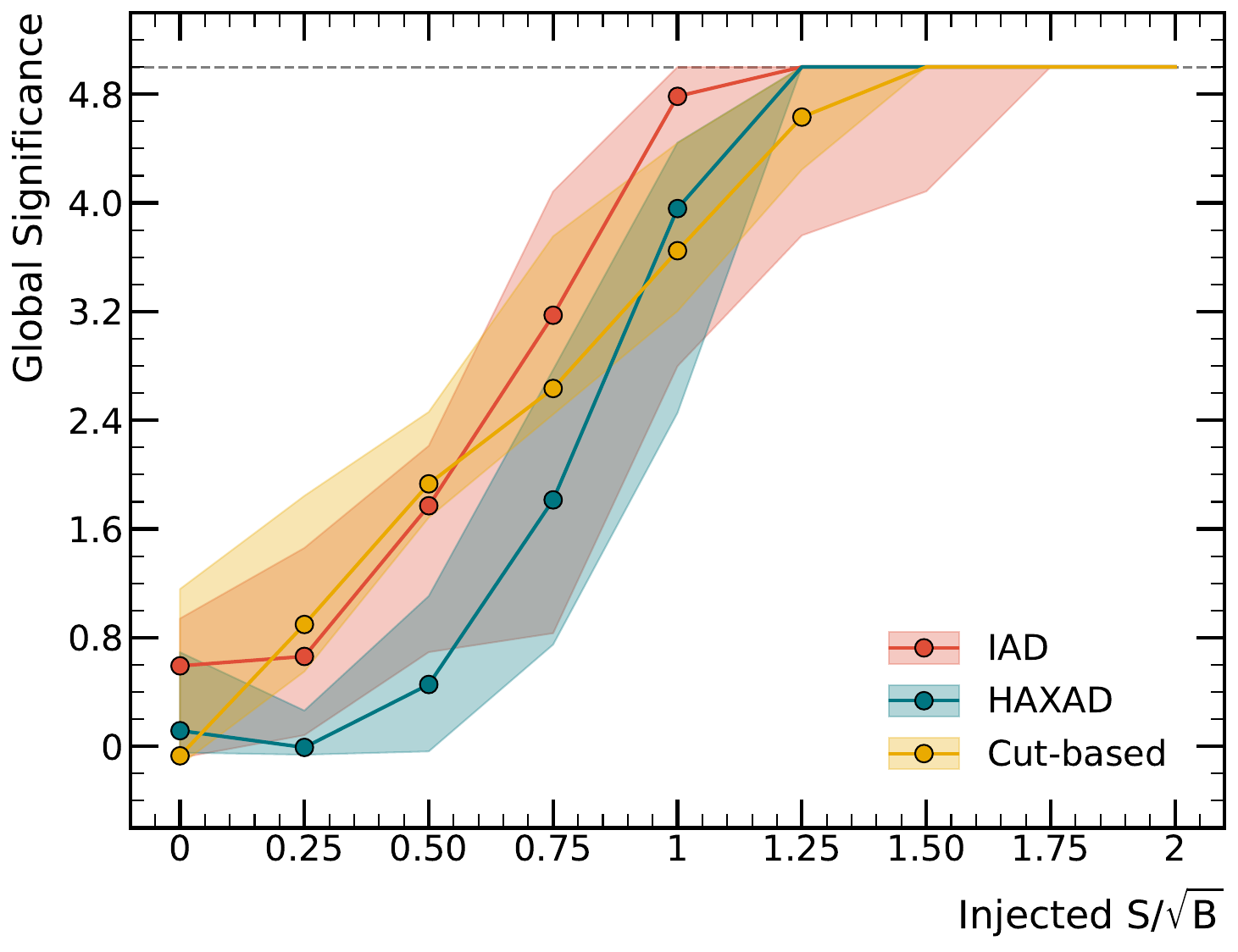}
    }%
    \hfill
    \subfloat[]{\label{fig:scan:XSH}
        \centering
        \includegraphics[width=0.45\linewidth]{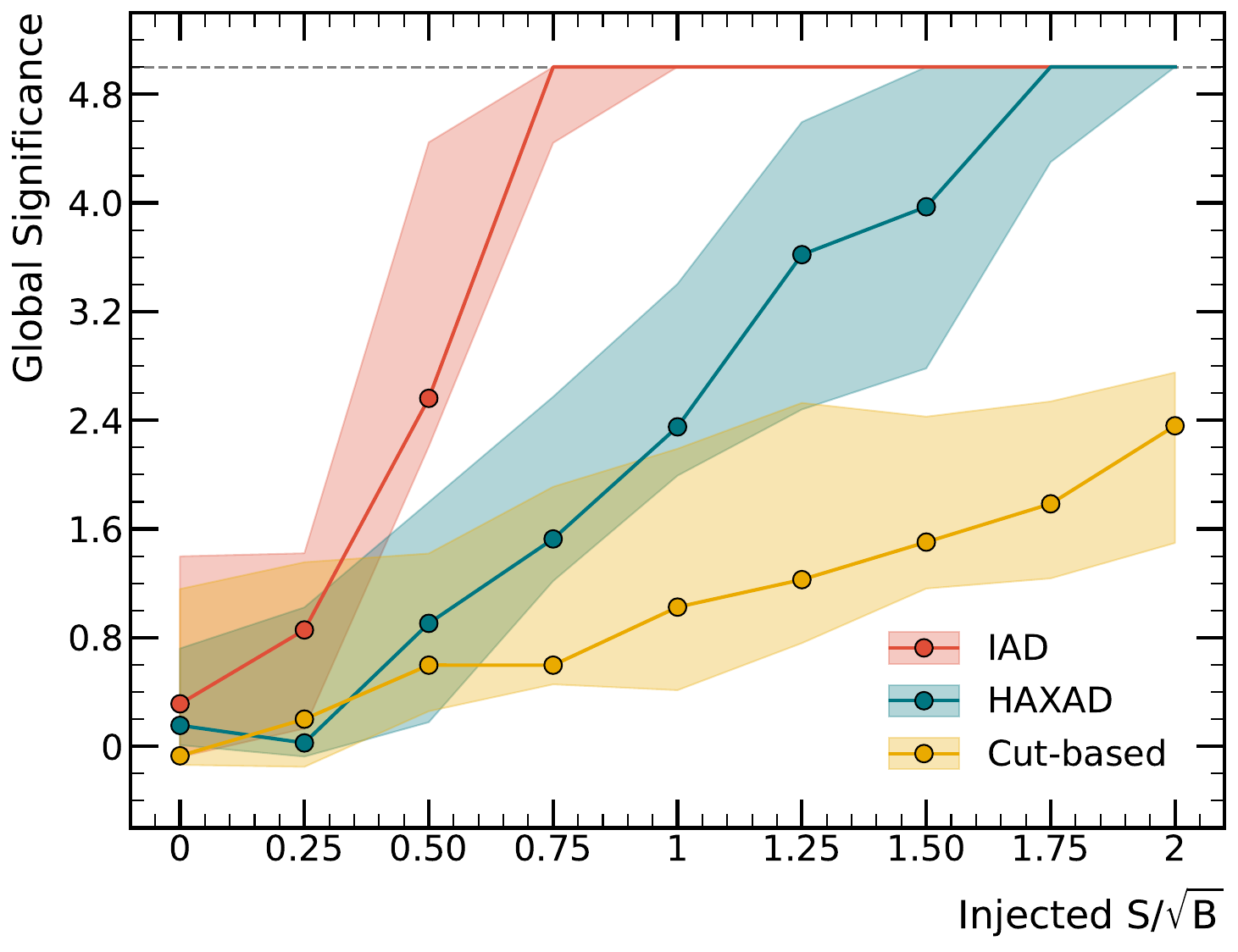}
    }
    \caption{
        Observed significances of the HAXAD method in comparison to the \gls{IAD} and the cut-based method, with $E_T^{\text{Miss}} > \SI{80}{\GeV}$, over different injected pre-selection significances.
        Shown are the results for a) the \SUSYSignal and b) the \XSHSignal benchmark signal.
    }
    \label{fig:scan}
\end{figure*}

The global significance values are scanned over different signal injection significances using the \SUSYSignal (\XSHSignal) benchmark signal are shown in \cref{fig:scan:SUSY150} (\cref{fig:scan:XSH}).
At each signal injection strength, 10 different random instances of the initial datasets are used to get the expected global significance with uncertainty.
The HAXAD method is compared with the \gls{IAD} method and the cut-based method, where the one-dimensional cut is optimized for the \SUSYSignal benchmark signal as $E_T^{\text{Miss}} > \SI{80}{\GeV}$.
The \gls{IAD} method matches or exceeds this optimized cut-based method for all values of signal injection.
The HAXAD method exceeds the optimized cut-based method for signal injections greater than 1\,$\sigma$.
The picture changes for the \XSHSignal benchmark signal for which the cut-based analysis is not optimized.
For this signal model, both the \gls{IAD} method and the HAXAD method exceed the cut-based method by a large amount for signal injections greater than 0.75\,$\sigma$.
For the considered amount of pseudo-experiments in the inference, the upper limit of the global significance is 5\,$\sigma$, and both HAXAD and the \gls{IAD} are reaching this limit at the 1.25\,$\sigma$ (1.75\,$\sigma$) injection of the \SUSYSignal (\XSHSignal) benchmark signal.

\section{Conclusion and Outlook}
\label{sec:conclusion}

Model-agnostic \gls{AD} methods provide an important complement to dedicated searches in high-energy physics. 
A wide range of \gls{AD} approaches exist; however, many struggle in the presence of rare processes or resonances that are already accounted for in the \gls{SM}, as the method can interpret the \gls{SM} signatures as potential anomalies.
This has limited the deployment of \gls{AD} searches in interesting regions, such as in proximity to the Higgs mass.

We introduced HAXAD, a method for incorporating limited information about the \gls{SM} into \gls{AD} searches, and demonstrated its use on \gls{BSM} signals associated with a \gls{SM} Higgs boson.
We achieved this by combining the data-driven background estimation of \gls{CATHODE}-like methods with simulations of \gls{SM} Higgs events.
We were able to demonstrate that a \gls{VAE}-based encoding can be used to overcome challenges in the generative background estimation, with a limited sacrifice to the sensitivity.
Our approach uses a large number of features from each event for the classification, and paves the way for future searches to use a feature set that is extended even beyond that, up to and including the use of low-level jet constituents as inputs.

While being fully model agnostic, the new method was benchmarked using two different \gls{BSM} signals.
When injecting the benchmark signals into the dataset with a a pre-selection significance of 1$\sigma$, our method was able to achieve a classification with a significance improvement of 5.4 (3.7) and a total significance of 4.0\,$\sigma$ (2.4\,$\sigma$) for the \SUSYSignal (\XSHSignal) benchmark signal, respectively.
For both benchmark signals, HAXAD remains competitive with, and outperforms, a dedicated cut-based search, while requiring no assumption about the respective signal.
As such, HAXAD provides a blueprint for future \gls{AD} searches around the Higgs peak or other resonances.

\section*{Code Availability}
The simulation framework, including all used configurations, is available at \url{https://github.com/hep-lbdl/EventGen}.
The analysis code is available at \url{https://gitlab.cern.ch/haxad/haxad_demonstrator}.

\section*{Acknowledgments}
C.L.C., S.Di., B.N., and D.N. are supported by the U.S. Department of Energy (DOE), Office of Science under contract DE-AC02-05CH11231, B.N. and D.N. are supported by DOE grant DE-AC02-76SF00515, S.D. and R.L. are supported by DOE grant DE-SC0017660.
B.N. is additionally supported by the John Templeton Foundation. 
C.L.C. is also supported by the U.S. DOE Office of Science under Contract No. DE-SC0017647.
This research used resources of the National Energy Research Scientific Computing Center (NERSC), a Department of Energy User Facility using NERSC award HEP-ERCAP0032304.

\bibliography{HEPML,other}

\providecommand{\href}[2]{#2}\begingroup\raggedright\begin{thebibliography}{10}

\bibitem{Bertone:2004pz}
G.~Bertone, D.~Hooper, and J.~Silk, {\it {Particle dark matter: Evidence, candidates and constraints}},  {\em Phys. Rept.} {\bf 405} (2005) 279--390, [\href{http://arxiv.org/abs/hep-ph/0404175}{{\tt hep-ph/0404175}}].

\bibitem{Arkani-Hamed:1998jmv}
N.~Arkani-Hamed, S.~Dimopoulos, and G.~R. Dvali, {\it {The Hierarchy problem and new dimensions at a millimeter}},  {\em Phys. Lett. B} {\bf 429} (1998) 263--272, [\href{http://arxiv.org/abs/hep-ph/9803315}{{\tt hep-ph/9803315}}].

\bibitem{Kasieczka:2021xcg}
G.~Kasieczka et~al., {\it {The LHC Olympics 2020 a community challenge for anomaly detection in high energy physics}},  {\em Rept. Prog. Phys.} {\bf 84} (2021), no.~12 124201, [\href{http://arxiv.org/abs/2101.08320}{{\tt arXiv:2101.08320}}].

\bibitem{Karagiorgi:2021ngt}
G.~Karagiorgi, G.~Kasieczka, S.~Kravitz, B.~Nachman, and D.~Shih, {\it {Machine Learning in the Search for New Fundamental Physics}},  {\em Nature Reviews Physics} {\bf 4} (12, 2021) 399--412, [\href{http://arxiv.org/abs/2112.03769}{{\tt arXiv:2112.03769}}].

\bibitem{Aarrestad:2021oeb}
T.~Aarrestad et~al., {\it {The Dark Machines Anomaly Score Challenge: Benchmark Data and Model Independent Event Classification for the Large Hadron Collider}},  {\em SciPost Phys.} {\bf 12} (2022), no.~1 043, [\href{http://arxiv.org/abs/2105.14027}{{\tt arXiv:2105.14027}}].

\bibitem{Belis:2023mqs}
V.~Belis, P.~Odagiu, and T.~K. Aarrestad, {\it {Machine learning for anomaly detection in particle physics}},  {\em Rev. Phys.} {\bf 12} (2024) 100091, [\href{http://arxiv.org/abs/2312.14190}{{\tt arXiv:2312.14190}}].

\bibitem{ATLAS:2023azi}
{\bf ATLAS} Collaboration, G.~Aad et~al., {\it {Anomaly detection search for new resonances decaying into a Higgs boson and a generic new particle $X$ in hadronic final states using $\sqrt{s} = 13$ TeV $pp$ collisions with the ATLAS detector}},  {\em Phys. Rev. D} {\bf 108} (2023) 052009, [\href{http://arxiv.org/abs/2306.03637}{{\tt arXiv:2306.03637}}].

\bibitem{ATLAS:2023ixc}
{\bf ATLAS} Collaboration, G.~Aad et~al., {\it {Search for New Phenomena in Two-Body Invariant Mass Distributions Using Unsupervised Machine Learning for Anomaly Detection at s=13{\,}{\,}TeV with the ATLAS Detector}},  {\em Phys. Rev. Lett.} {\bf 132} (2024), no.~8 081801, [\href{http://arxiv.org/abs/2307.01612}{{\tt arXiv:2307.01612}}].

\bibitem{ATLAS:2020iwa}
{\bf ATLAS} Collaboration, G.~Aad et~al., {\it {Dijet resonance search with weak supervision using $\sqrt{s}=13$ TeV $pp$ collisions in the ATLAS detector}},  {\em Phys. Rev. Lett.} {\bf 125} (2020), no.~13 131801, [\href{http://arxiv.org/abs/2005.02983}{{\tt arXiv:2005.02983}}].

\bibitem{ATLAS:2025obc}
{\bf ATLAS} Collaboration, G.~Aad et~al., {\it {Weakly supervised anomaly detection for resonant new physics in the dijet final state using proton-proton collisions at $\sqrt{s}=13$ TeV with the ATLAS detector}},  \href{http://arxiv.org/abs/2502.09770}{{\tt arXiv:2502.09770}}.

\bibitem{CMS:2024nsz}
{\bf CMS} Collaboration, V.~Chekhovsky et~al., {\it {Model-agnostic search for dijet resonances with anomalous jet substructure in proton{\textendash}proton collisions at $\sqrt{s}$ = 13 TeV}},  {\em Rept. Prog. Phys.} {\bf 88} (2025), no.~6 067802, [\href{http://arxiv.org/abs/2412.03747}{{\tt arXiv:2412.03747}}].

\bibitem{Metodiev:2017vrx}
E.~M. Metodiev, B.~Nachman, and J.~Thaler, {\it {Classification without labels: Learning from mixed samples in high energy physics}},  {\em JHEP} {\bf 10} (2017) 174, [\href{http://arxiv.org/abs/1708.02949}{{\tt arXiv:1708.02949}}].

\bibitem{Collins:2018epr}
J.~H. Collins, K.~Howe, and B.~Nachman, {\it {Anomaly Detection for Resonant New Physics with Machine Learning}},  {\em Phys. Rev. Lett.} {\bf 121} (2018), no.~24 241803, [\href{http://arxiv.org/abs/1805.02664}{{\tt arXiv:1805.02664}}].

\bibitem{Collins:2019jip}
J.~H. Collins, K.~Howe, and B.~Nachman, {\it {Extending the search for new resonances with machine learning}},  {\em Phys. Rev. D} {\bf 99} (2019), no.~1 014038, [\href{http://arxiv.org/abs/1902.02634}{{\tt arXiv:1902.02634}}].

\bibitem{Hallin:2021wme}
A.~Hallin, J.~Isaacson, G.~Kasieczka, C.~Krause, B.~Nachman, T.~Quadfasel, M.~Schlaffer, D.~Shih, and M.~Sommerhalder, {\it {Classifying anomalies through outer density estimation}},  {\em Phys. Rev. D} {\bf 106} (2022), no.~5 055006, [\href{http://arxiv.org/abs/2109.00546}{{\tt arXiv:2109.00546}}].

\bibitem{Golfand:1971iw}
Y.~A. Golfand and E.~P. Likhtman, {\it {Extension of the Algebra of Poincare Group Generators and Violation of p Invariance}},  {\em JETP Lett.} {\bf 13} (1971) 323--326.

\bibitem{Volkov:1973ix}
D.~V. Volkov and V.~P. Akulov, {\it {Is the Neutrino a Goldstone Particle?}},  {\em Phys. Lett. B} {\bf 46} (1973) 109--110.

\bibitem{Wess:1974tw}
J.~Wess and B.~Zumino, {\it {Supergauge Transformations in Four-Dimensions}},  {\em Nucl. Phys. B} {\bf 70} (1974) 39--50.

\bibitem{Wess:1974jb}
J.~Wess and B.~Zumino, {\it {Supergauge Invariant Extension of Quantum Electrodynamics}},  {\em Nucl. Phys. B} {\bf 78} (1974) 1.

\bibitem{Ferrara:1974pu}
S.~Ferrara and B.~Zumino, {\it {Supergauge Invariant Yang-Mills Theories}},  {\em Nucl. Phys. B} {\bf 79} (1974) 413.

\bibitem{Salam:1974ig}
A.~Salam and J.~A. Strathdee, {\it {Supersymmetry and Nonabelian Gauges}},  {\em Phys. Lett. B} {\bf 51} (1974) 353--355.

\bibitem{Guasch:1999jp}
J.~Guasch and J.~Sola, {\it {FCNC top quark decays: A Door to SUSY physics in high luminosity colliders?}},  {\em Nucl. Phys. B} {\bf 562} (1999) 3--28, [\href{http://arxiv.org/abs/hep-ph/9906268}{{\tt hep-ph/9906268}}].

\bibitem{Bejar:2000ub}
S.~Bejar, J.~Guasch, and J.~Sola, {\it {Loop induced flavor changing neutral decays of the top quark in a general two Higgs doublet model}},  {\em Nucl. Phys. B} {\bf 600} (2001) 21--38, [\href{http://arxiv.org/abs/hep-ph/0011091}{{\tt hep-ph/0011091}}].

\bibitem{Eilam:2001dh}
G.~Eilam, A.~Gemintern, T.~Han, J.~M. Yang, and X.~Zhang, {\it {Top quark rare decay t ---{\ensuremath{>}} ch in R-parity violating SUSY}},  {\em Phys. Lett. B} {\bf 510} (2001) 227--235, [\href{http://arxiv.org/abs/hep-ph/0102037}{{\tt hep-ph/0102037}}].

\bibitem{Aguilar-Saavedra:2002phh}
J.~A. Aguilar-Saavedra, {\it {Effects of mixing with quark singlets}},  {\em Phys. Rev. D} {\bf 67} (2003) 035003, [\href{http://arxiv.org/abs/hep-ph/0210112}{{\tt hep-ph/0210112}}]. [Erratum: Phys.Rev.D 69, 099901 (2004)].

\bibitem{Cao:2007dk}
J.~J. Cao, G.~Eilam, M.~Frank, K.~Hikasa, G.~L. Liu, I.~Turan, and J.~M. Yang, {\it {SUSY-induced FCNC top-quark processes at the large hadron collider}},  {\em Phys. Rev. D} {\bf 75} (2007) 075021, [\href{http://arxiv.org/abs/hep-ph/0702264}{{\tt hep-ph/0702264}}].

\bibitem{delAguila:1982fs}
F.~del Aguila and M.~J. Bowick, {\it {The Possibility of New Fermions With $\Delta$ I = 0 Mass}},  {\em Nucl. Phys. B} {\bf 224} (1983) 107.

\bibitem{Aguilar-Saavedra:2009xmz}
J.~A. Aguilar-Saavedra, {\it {Identifying top partners at LHC}},  {\em JHEP} {\bf 11} (2009) 030, [\href{http://arxiv.org/abs/0907.3155}{{\tt arXiv:0907.3155}}].

\bibitem{ATLAS:2023omk}
{\bf ATLAS} Collaboration, G.~Aad et~al., {\it {Model-independent search for the presence of new physics in events including $H\rightarrow\gamma\gamma$ with $\sqrt{s}$ = 13 TeV pp data recorded by the ATLAS detector at the LHC}},  {\em JHEP} {\bf 07} (2023) 176, [\href{http://arxiv.org/abs/2301.10486}{{\tt arXiv:2301.10486}}].

\bibitem{ATLAS:2008xda}
{\bf ATLAS} Collaboration, G.~Aad et~al., {\it {The ATLAS Experiment at the CERN Large Hadron Collider}},  {\em JINST} {\bf 3} (2008) S08003.

\bibitem{CMS:2008xjf}
{\bf CMS} Collaboration, C.~Collaboration, {\it {The CMS Experiment at the CERN LHC}},  {\em JINST} {\bf 3} (2008) S08004.

\bibitem{ATLAS:2018zdn}
{\bf ATLAS} Collaboration, M.~Aaboud et~al., {\it {A strategy for a general search for new phenomena using data-derived signal regions and its application within the ATLAS experiment}},  {\em Eur. Phys. J. C} {\bf 79} (2019), no.~2 120, [\href{http://arxiv.org/abs/1807.07447}{{\tt arXiv:1807.07447}}].

\bibitem{CMS:2020zjg}
{\bf CMS} Collaboration, A.~M. Sirunyan et~al., {\it {MUSiC: a model-unspecific search for new physics in proton{\textendash}proton collisions at $\sqrt{s} = 13\,\text {TeV} $}},  {\em Eur. Phys. J. C} {\bf 81} (2021), no.~7 629, [\href{http://arxiv.org/abs/2010.02984}{{\tt arXiv:2010.02984}}].

\bibitem{Alwall:2011uj}
J.~Alwall, M.~Herquet, F.~Maltoni, O.~Mattelaer, and T.~Stelzer, {\it {MadGraph 5 : Going Beyond}},  {\em JHEP} {\bf 06} (2011) 128, [\href{http://arxiv.org/abs/1106.0522}{{\tt arXiv:1106.0522}}].

\bibitem{Sjostrand:2006za}
T.~Sjostrand, S.~Mrenna, and P.~Z. Skands, {\it {PYTHIA 6.4 Physics and Manual}},  {\em JHEP} {\bf 05} (2006) 026, [\href{http://arxiv.org/abs/hep-ph/0603175}{{\tt hep-ph/0603175}}].

\bibitem{Sjostrand:2014zea}
T.~Sj{\"o}strand, S.~Ask, J.~R. Christiansen, R.~Corke, N.~Desai, P.~Ilten, S.~Mrenna, S.~Prestel, C.~O. Rasmussen, and P.~Z. Skands, {\it {An introduction to PYTHIA 8.2}},  {\em Comput. Phys. Commun.} {\bf 191} (2015) 159--177, [\href{http://arxiv.org/abs/1410.3012}{{\tt arXiv:1410.3012}}].

\bibitem{deFavereau:2013fsa}
{\bf DELPHES 3} Collaboration, J.~de~Favereau, C.~Delaere, P.~Demin, A.~Giammanco, V.~Lema{\^\i}tre, A.~Mertens, and M.~Selvaggi, {\it {DELPHES 3, A modular framework for fast simulation of a generic collider experiment}},  {\em JHEP} {\bf 02} (2014) 057, [\href{http://arxiv.org/abs/1307.6346}{{\tt arXiv:1307.6346}}].

\bibitem{Mertens:2015kba}
A.~Mertens, {\it {New features in Delphes 3}},  {\em J. Phys. Conf. Ser.} {\bf 608} (2015), no.~1 012045.

\bibitem{Fayet:1976et}
P.~Fayet, {\it {Supersymmetry and Weak, Electromagnetic and Strong Interactions}},  {\em Phys. Lett. B} {\bf 64} (1976) 159.

\bibitem{Fayet:1977yc}
P.~Fayet, {\it {Spontaneously Broken Supersymmetric Theories of Weak, Electromagnetic and Strong Interactions}},  {\em Phys. Lett. B} {\bf 69} (1977) 489.

\bibitem{Farrar:1978xj}
G.~R. Farrar and P.~Fayet, {\it {Phenomenology of the Production, Decay, and Detection of New Hadronic States Associated with Supersymmetry}},  {\em Phys. Lett. B} {\bf 76} (1978) 575--579.

\bibitem{Alwall:2008ag}
J.~Alwall, P.~Schuster, and N.~Toro, {\it {Simplified Models for a First Characterization of New Physics at the LHC}},  {\em Phys. Rev. D} {\bf 79} (2009) 075020, [\href{http://arxiv.org/abs/0810.3921}{{\tt arXiv:0810.3921}}].

\bibitem{LHCNewPhysicsWorkingGroup:2011mji}
{\bf LHC New Physics Working Group} Collaboration, D.~Alves, {\it {Simplified Models for LHC New Physics Searches}},  {\em J. Phys. G} {\bf 39} (2012) 105005, [\href{http://arxiv.org/abs/1105.2838}{{\tt arXiv:1105.2838}}].

\bibitem{ATLAS:2020qlk}
{\bf ATLAS} Collaboration, G.~Aad et~al., {\it {Search for direct production of electroweakinos in final states with missing transverse momentum and a Higgs boson decaying into photons in pp collisions at $ \sqrt{s} $ = 13 TeV with the ATLAS detector}},  {\em JHEP} {\bf 10} (2020) 005, [\href{http://arxiv.org/abs/2004.10894}{{\tt arXiv:2004.10894}}].

\bibitem{Robens:2019kga}
T.~Robens, T.~Stefaniak, and J.~Wittbrodt, {\it {Two-real-scalar-singlet extension of the SM: LHC phenomenology and benchmark scenarios}},  {\em Eur. Phys. J. C} {\bf 80} (2020), no.~2 151, [\href{http://arxiv.org/abs/1908.08554}{{\tt arXiv:1908.08554}}].

\bibitem{Basler:2018dac}
P.~Basler, S.~Dawson, C.~Englert, and M.~M{\"u}hlleitner, {\it {Showcasing HH production: Benchmarks for the LHC and HL-LHC}},  {\em Phys. Rev. D} {\bf 99} (2019), no.~5 055048, [\href{http://arxiv.org/abs/1812.03542}{{\tt arXiv:1812.03542}}].

\bibitem{Baum:2019pqc}
S.~Baum and N.~R. Shah, {\it {Benchmark Suggestions for Resonant Double Higgs Production at the LHC for Extended Higgs Sectors}},  \href{http://arxiv.org/abs/1904.10810}{{\tt arXiv:1904.10810}}.

\bibitem{CMS:2022suh}
{\bf CMS} Collaboration, A.~Tumasyan et~al., {\it {Search for a massive scalar resonance decaying to a light scalar and a Higgs boson in the four b quarks final state with boosted topology}},  {\em Phys. Lett. B} {\bf 842} (2023) 137392, [\href{http://arxiv.org/abs/2204.12413}{{\tt arXiv:2204.12413}}].

\bibitem{CMS:2021yci}
{\bf CMS} Collaboration, A.~Tumasyan et~al., {\it {Search for a heavy Higgs boson decaying into two lighter Higgs bosons in the $\tau\tau$bb final state at 13 TeV}},  {\em JHEP} {\bf 11} (2021) 057, [\href{http://arxiv.org/abs/2106.10361}{{\tt arXiv:2106.10361}}].

\bibitem{CMS:2023boe}
{\bf CMS} Collaboration, A.~Tumasyan et~al., {\it {Search for a new resonance decaying into two spin-0 bosons in a final state with two photons and two bottom quarks in proton-proton collisions at $ \sqrt{s} $ = 13 TeV}},  {\em JHEP} {\bf 05} (2024) 316, [\href{http://arxiv.org/abs/2310.01643}{{\tt arXiv:2310.01643}}].

\bibitem{ATLAS:2023tkl}
{\bf ATLAS} Collaboration, G.~Aad et~al., {\it {Search for a new heavy scalar particle decaying into a Higgs boson and a new scalar singlet in final states with one or two light leptons and a pair of {\ensuremath{\tau}}-leptons with the ATLAS detector}},  {\em JHEP} {\bf 10} (2023) 009, [\href{http://arxiv.org/abs/2307.11120}{{\tt arXiv:2307.11120}}].

\bibitem{Cacciari:2005hq}
M.~Cacciari and G.~P. Salam, {\it {Dispelling the $N^{3}$ myth for the $k_t$ jet-finder}},  {\em Phys. Lett. B} {\bf 641} (2006) 57--61, [\href{http://arxiv.org/abs/hep-ph/0512210}{{\tt hep-ph/0512210}}].

\bibitem{Cacciari:2011ma}
M.~Cacciari, G.~P. Salam, and G.~Soyez, {\it {FastJet User Manual}},  {\em Eur. Phys. J. C} {\bf 72} (2012) 1896, [\href{http://arxiv.org/abs/1111.6097}{{\tt arXiv:1111.6097}}].

\bibitem{Cacciari:2008gp}
M.~Cacciari, G.~P. Salam, and G.~Soyez, {\it {The anti-$k_t$ jet clustering algorithm}},  {\em JHEP} {\bf 04} (2008) 063, [\href{http://arxiv.org/abs/0802.1189}{{\tt arXiv:0802.1189}}].

\bibitem{kingma2014autoencoding}
D.~P. Kingma and M.~Welling, {\it Auto-encoding variational bayes},  \href{http://arxiv.org/abs/1312.6114}{{\tt arXiv:1312.6114}}.

\bibitem{DBLP:journals/corr/abs-1912-01703}
A.~Paszke et~al., {\it Pytorch: An imperative style, high-performance deep learning library},  {\em CoRR} {\bf abs/1912.01703} (2019) [\href{http://arxiv.org/abs/1912.01703}{{\tt arXiv:1912.01703}}].

\bibitem{adam}
D.~Kingma and J.~Ba, {\it Adam: A method for stochastic optimization},  \href{http://arxiv.org/abs/1412.6980}{{\tt arXiv:1412.6980}}.

\bibitem{Hallin:2022eoq}
A.~Hallin, G.~Kasieczka, T.~Quadfasel, D.~Shih, and M.~Sommerhalder, {\it {Resonant anomaly detection without background sculpting}},  {\em Phys. Rev. D} {\bf 107} (2023), no.~11 114012, [\href{http://arxiv.org/abs/2210.14924}{{\tt arXiv:2210.14924}}].

\bibitem{papamakarios2021normalizingflowsprobabilisticmodeling}
G.~Papamakarios, E.~Nalisnick, D.~J. Rezende, S.~Mohamed, and B.~Lakshminarayanan, {\it Normalizing flows for probabilistic modeling and inference},  {\em Journal of Machine Learning Research} {\bf 22} (2021), no.~57 1--64, [\href{http://arxiv.org/abs/1912.02762}{{\tt arXiv:1912.02762}}].

\bibitem{nflows}
C.~Durkan, A.~Bekasov, I.~Murray, and G.~Papamakarios, {\it {nflows}: normalizing flows in {PyTorch}},  Nov., 2020.

\bibitem{durkan2019neural}
C.~Durkan, A.~Bekasov, I.~Murray, and G.~Papamakarios, {\it Neural spline flows},  {\em Advances in neural information processing systems} {\bf 32} (2019) [\href{http://arxiv.org/abs/1906.04032}{{\tt arXiv:1906.04032}}].

\bibitem{lipman2023flowmatchinggenerativemodeling}
Y.~Lipman, R.~T.~Q. Chen, H.~Ben-Hamu, M.~Nickel, and M.~Le, {\it Flow matching for generative modeling},  \href{http://arxiv.org/abs/2210.02747}{{\tt arXiv:2210.02747}}.

\bibitem{Diefenbacher:2020rna}
S.~Diefenbacher, E.~Eren, G.~Kasieczka, A.~Korol, B.~Nachman, and D.~Shih, {\it {DCTRGAN: Improving the Precision of Generative Models with Reweighting}},  {\em JINST} {\bf 15} (2020), no.~11 P11004, [\href{http://arxiv.org/abs/2009.03796}{{\tt arXiv:2009.03796}}].

\bibitem{Finke:2023ltw}
T.~Finke, M.~Hein, G.~Kasieczka, M.~Kr{\"a}mer, A.~M{\"u}ck, P.~Prangchaikul, T.~Quadfasel, D.~Shih, and M.~Sommerhalder, {\it {Tree-based algorithms for weakly supervised anomaly detection}},  {\em Phys. Rev. D} {\bf 109} (2024), no.~3 034033, [\href{http://arxiv.org/abs/2309.13111}{{\tt arXiv:2309.13111}}].

\bibitem{Freytsis:2023cjr}
M.~Freytsis, M.~Perelstein, and Y.~C. San, {\it {Anomaly detection in the presence of irrelevant features}},  {\em JHEP} {\bf 02} (2024) 220, [\href{http://arxiv.org/abs/2310.13057}{{\tt arXiv:2310.13057}}].

\bibitem{DBLP:journals/corr/ChenG16}
T.~Chen and C.~Guestrin, {\it Xgboost: {A} scalable tree boosting system},  {\em CoRR} {\bf abs/1603.02754} (2016) [\href{http://arxiv.org/abs/1603.02754}{{\tt arXiv:1603.02754}}].

\bibitem{Vaslin:2022bds}
L.~Vaslin, S.~Calvet, V.~Barra, and J.~Donini, {\it {pyBumpHunter: A model independent bump hunting tool in Python for High Energy Physics analyses}},  {\em SciPost Phys. Codeb.} {\bf 2023} (2023) 15, [\href{http://arxiv.org/abs/2208.14760}{{\tt arXiv:2208.14760}}].

\end{thebibliography}\endgroup
\bibliographystyle{JHEP}

\clearpage

\appendix

\end{document}